\def\dirac#1{\mathord{\not\mathrel{\mskip 2mu #1}}}
\documentstyle[12pt]{article}

\def\beqra{\begin{eqnarray}}
\def\eeqra{\end{eqnarray}}
\def\beqast{\begin{eqnarray*}}
\def\eeqast{\end{eqnarray*}}
\def\be{\begin{enumerate}}
\def\ee{\end{enumerate}}
\def\lag{\langle}
\def\rag{\rangle}

\def\sppt{Research supported in part by the
Robert A. Welch Foundation and NSF Grant PHY 9009850}

\def\twohead#1{\vskip\baselineskip\noindent{\normalsize\it #1}\vskip8pt}
\def\utgp{Theory Group\\ Department of Physics \\ University of Texas
\\ Austin, Texas 78712}
\def\utgp{Theory Group\\ Department of Physics \\ University of Texas
\\ Austin, Texas 78712}

\def\fnote#1#2{\begingroup\def\thefootnote{#1}\footnote{#2}\addtocounter
{footnote}{-1}\endgroup}

\def\beq{\begin{equation}}	\def\eeq{\end{equation}}
\def\haf{\frac{1}{2}}
\def\pa{\partial}

\begin{document}

\hfill{UTTG-25-92}

\vspace{24pt}

\begin{center}
{\bf   Conference Summary}
\vspace{36pt}

 Steven Weinberg\fnote{*}{\sppt.}

\vspace{24pt}
\utgp

\vspace{30pt}

\begin{minipage}{4.75in}
Talk presented at the XXVI International Conference on High\\ Energy Physics,
Dallas, Texas, August, 12, 1992.

\end{minipage}

\vfill
\end{center}
\baselineskip=24pt
\pagebreak
\setcounter{page}{1}

Roy Schwitters has thanked the staff of this conference for their good work.
As this is the last talk, I would like to add my own thanks to the staff, and
also on behalf of all the participants to thank Roy himself and Vic Teplitz
and the other physicists at the
Super Collider and at SMU and the other Dallas area universities for their part
in making this such an exceptionally well run conference.

This is actually the second time that I have been called on to give the
summary talk at one of the Rochester Conferences.  The previous time was at
Berkeley in 1986, the latest time the conference was held in the
United States.  You might think that after that experience I would know better
than to take on such a difficult task, but in fact nothing could be further
from the truth.  This is actually an extremely easy job.  Everyone knows that
it is impossible to review every topic that has been discussed at such a
conference, so no one expects it.  In fact I found last time
at Berkeley that people (at least most
people) even forgave me for not mentioning their own work.

It {\em is} possible to give a brief ``coarse-grained"
summary of the whole conference.  This also is very easy, because it doesn't
vary from one conference to another.  Here it is:
\begin{enumerate}
\item
The Standard Model agrees with all data, but has many holes and loose ends.
\item
We must still:
\begin{itemize}
\item
find the top quark,
\item
identify the mechanism (or mechanisms) for CP nonconservation,
\item
solve quantum chromodynamics.
\end{itemize}
\item
We don't understand:
\begin{itemize}
\item
why the parameters of the standard model take the values we observe,
\item
why there are three generations of quarks and leptons,
\item
why $SU(3)\otimes SU(2)\otimes U(1)$?
\end{itemize}
\item
Most of these questions revolve around the central unresolved issue
concerning the Standard Model: how is $SU(2)\otimes U(1)$ broken?  There are
two
possibilities: The Goldstone bosons which provide the longitudinal parts of the
W
and the Z are either elementary, or they are not.
\begin{itemize}
\item
  If the Goldstone bosons are part of a multiplet of elementary scalars, then
in order to explain why they are so light compared to a really fundamental
scale we probably need to assume supersymmetry.
\item
If the Goldstone bosons are composites, then definitely
there must be new extra-strong forces, like the technicolor forces reviewed
here by Appelquist. \end{itemize}
(Peskin in his talk referred to these alternatives
 as the weak-interaction route and the strong-interaction route.)
\item
In order to resolve the issue of electroweak symmetry breaking and break out of
the present impasse, we need the
\begin{itemize}
\item
SSC
\item
LHC.
\end{itemize}
\end{enumerate}
\begin{center}
*\qquad *\qquad *
\end{center}

In the remaining 57 minutes of my talk I would like to discuss a few specific
topics.  These are chosen, not necessarily because they represent the most
important things discussed at this conference, but mostly because they are
matters that I found interesting and about which I had something that I wanted
to say. In
1986 the two topics I chose to discuss were solar neutrino oscillations and
string
theory.  (By the way,  Robertson\footnote{I will follow the practice here for
the most part of just quoting talks given at this conference, chiefly the
rapporteur talks.  References to the physics literature can be found in the
written versions of these talks.} said in his talk the other day that only
recently
have particle physicists regarded the solar neutrino problem as part of
particle physics, which is true only if ``recently" extends back at least six
years.)  These would be a pretty good choice of a pair of special topics for my
talk today (which shows how slowly our field is moving), and I will come back
to them, but there are a few other topics I want to take up here also.  At the
end of this discussion of special topics I will say a little about what
I think lies ahead for particle physics.

\twohead{Heavy Quark Symmetry}

My first topic is heavy quark symmetry.  This was discussed by Drell and Isgur
in their rapporteur's talks; summaries were given in parallel sessions by Close
and Grinstein; and the subject  kept coming up in many of the
talks
in parallel sessions I attended.  It is really one of the prettiest
developments in the theory of strong interactions in a long time,  and one in
that in a remarkable way has almost
immediately been  taken over by experimentalists as part of their standard bag
of tricks.  Heavy quark symmetry has been explained many times in this
conference in a hand-waving way in terms of analogies with atomic physics.
Here I would like to
offer an explanation (most of which can be found in the papers of the experts),
that you can give with your hands strapped to your sides,
in terms of the Feynman diagrams of quantum chromodynamics.

Consider a heavy quark line going through a Feynman diagram.  Suppose that
after emitting momentum to and absorbing momentum from the gluons, it has
acquired a four-momentum $mv+k$, where $m$ and $v$ are the mass and initial
four-velocity ($v\equiv[\gamma\vec{v},\gamma]$) of the heavy quark.  The
components of the four-momentum transfer are limited by
an amount of the order of the QCD scale factor $\Lambda$, so for $m\gg
\Lambda$, the quark propagator may be approximated by
\beq
\frac{-i(m\dirac{v}+\dirac{k})+m}{(mv+k)^2+m^2}\simeq
\frac{-i\dirac{v}+1}{2v\cdot k}\;.
\eeq
[I am using the usual slash notation, $\dirac{a}\equiv \gamma^\mu a_\mu$, with
a metric that has diagonal components $-1,+1,+1,+1$.]  This immediately reveals
the flavor degeneracy of these theories; as long as we express everything in
terms of velocities rather than four-momenta, nothing in matrix elements
depends on the heavy quark mass.
With $N$ heavy types of heavy quarks, the bound states of one or more heavy
quarks plus any number of light quarks and gluons would form $SU(N)$ multiplets
of hadrons with equal {\em binding} energies.

To go further, suppose gluon lines with polarization indices $\mu$, $\nu$, etc.
and color indices $a$, $b$, etc., are emitted by a heavy quark line before it
finally leaves the diagram with spin $z$-component $\sigma$ and four-velocity
$v$.  By moving all factors $(-i\dirac{v}+1)/2$ to the end of the line, we
easily see that the contribution of the quark-gluon vertices and quark
propagators is
\beqra
&&\bar{u}(\sigma,v)\left(\frac{-i\dirac{v}+1}{2}\right)it_a \gamma^\mu
\left(\frac{-i\dirac{v}+1}{2}\right)\nonumber\\&& \times\; it_b \gamma^\nu
\left(\frac{-i\dirac{v}+1}{2}\right)\cdots
\\ &&\nonumber\qquad=
\bar{u}(\sigma,v)\left(\frac{-i\dirac{v}+1}{2}\right)\,t_a t_b\cdots v^\mu
v^\nu\cdots\;.
\eeqra
In other words, nothing depends on the spin $z$-components of the heavy quarks,
aside from kinematic final (and initial) state factors like \linebreak
$\bar{u}(\sigma,v)(-i\dirac{v}+1)/2$, which do not vary from diagram to diagram
for any given process.  (Despite appearances, this result does not depend on
the fact that the quarks have spin $\haf$.)  In particular, the
positions of the poles in a Feynman diagram will not depend on the spin
$z$-components of any incoming or outgoing heavy quarks, so the bound states of
heavy and light quarks will exhibit a spin degeneracy as well as a flavor
degeneracy: all of the bound states that are related by rotating only heavy
quark spins (as for instance the
lowest $0^-$ and $1^-$ bound states of a heavy quark and a light antiquark)
will have the same mass.

One can also apply this diagrammatic analysis to processes in which a weak
interaction induces a transition between hadrons containing different species
of heavy quarks.    According to the foregoing analysis, the matrix element of
a current $\bar{\psi}_{f'}\Gamma\psi_f$ (with $\Gamma$ an arbitrary $4\times 4$
matrix) between initial and
final bound states $\alpha$ and $\alpha'$, consisting of any number of light
quarks and/or antiquarks plus one heavy quark respectively of flavor $f$ and
$f'$ and four-velocity $v$ and $v'$,
must take the form:
\vadjust{\penalty -500}
\beqra
\lag f',v',\alpha'|\left(\bar{\psi}_{f'}\Gamma\psi_f\right)|f,v,\alpha\rag
  \span\span\hfill\nonumber\\
\noalign{\penalty 9999}
{}={}
&&\mskip -27mu
    \sum_{\sigma',\sigma}
    C_{\alpha\rightarrow\alpha'}(v',\sigma',v,\sigma) \\
\noalign{\penalty 9999}
&& \times\left[\bar{u}(\sigma',v')
    \textstyle{\left(\frac{-i\dirac{v'}+1}{2}\right)}\,
    \Gamma\,\textstyle{\left(\frac{-i\dirac{v}+1}{2}\right)}\,
    u(\sigma,v)\right] \;,\nonumber
\eeqra
where $C_{\alpha\rightarrow\alpha'}$ is an unknown but flavor-independent
function of initial and final heavy quark spin $z$-components and velocities,
that arises from the convolution of the wave functions for states $\alpha$ and
$\alpha'$ with the sum of Feynman diagrams in which the heavy quark line is
replaced by a product of factors
$v^\mu t_a$ for each interaction of gluons with the heavy quark.  (These
Feynman diagrams could actually be summed explicitly for Abelian gluons, but as
far as is known not for real SU(3) gluons.)  We can usefully rewrite this
formula as a trace:
\beqra
\lag f',v',\alpha'|\left(\bar{\psi}_{f'}\Gamma\psi_f\right)|f, v, \alpha\rag\\
\span\span
={\rm Tr}\left\{M_{\alpha\rightarrow\alpha'}(v',v)\,
\textstyle{\left(\frac{-i\dirac{v'}+1}{2}\right)}\,\Gamma\,
    \textstyle{\left(\frac{-i\dirac{v}+1}{2}\right)}\,\right\}\;,
\nonumber
\eeqra
where $M$ is the $4\times 4$ matrix:
\beqra
M_{\alpha\rightarrow\alpha'}(v',v)\\
\span\span\hfill
\equiv\sum_{\sigma,\sigma'}\, C_{\alpha\rightarrow\alpha'}(v',\sigma',v,\sigma)
    \, u(\sigma,v)\,\bar{u}(\sigma',v')\;.\nonumber
\eeqra
Though $M$ is unknown, its matrix structure is fixed by Lorentz invariance.
For instance, for a transition between the lowest $0^-$ mesons with flavor $f$
and $f'$, the above matrix element must have the same Lorentz transformation
properties as the matrix $\Gamma$, so $M_{0^-\rightarrow 0^-}(v',v)$ must be a
scalar.  Factors of $\dirac{v}$ and $\dirac{v
}'$ do not matter because when multiplied into $(-i\dirac{v}+1)$ or
$(-i\dirac{v}'+1)$ they merely yield factors of $+i$.  Thus $M$ must here be
proportional to the unit matrix:
\beq
M_{0^-\rightarrow 0^-}(v',v)=\haf\,\xi(-v\cdot v')\,1\;,
\eeq
with a coefficient $\xi$ that depends on the only scalar variable, $v\cdot v'$.
 This is the celebrated Isgur-Wise function.  If you like analogies, then think
of $\xi(-v\cdot v')$ as analogous to the well-known Fermi function $F(Z, W)$,
which gives the effect of final state Coulomb interactions in nuclear beta
decay.  The fact that
$\xi(-v\cdot v')$ depends neither on heavy quark flavor nor on the matrix
$\Gamma$ is just like the fact that  $F(Z, W)$ does not depend on the nature of
the nuclei participating in the beta transition [aside from the energy $W$,
which is analogous to the variable $v\cdot v'$, and the atomic number $Z$,
which is analogous to the fixed color triplet assignment of quarks], or on
the matrices [S, V, T, A, or P] appearing in the beta decay Hamiltonian.

It is a challenge to quantum chromodynamics to calculate the Isgur-Wise
function, and much effort has been put into this problem, but there is one
point where the value of $\xi$ can be obtained without effort.  In the special
case where $\Gamma=i\gamma^\mu$, $f=f'$, and $v=v'$, the matrix element (4) is
given by the conservation of heavy quark flavor as just $v^\mu$ [in much the
same way that the matrix
element for ${\rm O}^{14}$ beta decay is fixed by the conservation of the
isospin vector current], so in this case $\xi=1$ at the point $v\cdot v'=-1$.
But $\xi$ is a universal function, independent of $\Gamma$ and heavy quark
flavors, so in general
\beq
\xi(1)=1\;.
\eeq
This result is used in determinations of CKM matrix elements like $V_{cb}$ by
extrapolating data on weak processes such as $B\rightarrow D+\ell+\nu$ to the
point $v\cdot v'=-1$.

The same Isgur-Wise function enters into the matrix elements for transitions
involving the $1^-$ mesons that are degenerate with the lightest $0^-$ mesons.
Whatever the wave function for the lowest $0^-$ mesons, the heavy quark spin
degeneracy tells us that the wave function for
the degenerate $1^-$ meson [in its rest frame] with $J_z=0$ is given by
inserting an extra factor $(-1)^{2\sigma}$.  We must therefore take
$M_{0^-\rightarrow 1^-}$ as a scalar matrix function of the final vector meson
polarization $e^\mu$ that in the rest frame of the final heavy quark for vector
meson polarization vector $e^0=e^1=e^2=0,\, e^3=1$ gives a factor $\haf
(-)^{2\sigma'}\xi$ when acting to the left on $\bar{u}(v',\sigma')$.  Aside
from inconsequential terms involving
$\dirac{v}$ and/or $\dirac{v}'$, the unique matrix satisfying these
requirements is
\beq
 M_{0^-\rightarrow 1^-,e}(v',v)=\frac{i}{2}\;\xi(-v\cdot v')\, \gamma_5
\dirac{e}\;.
\eeq
This can be used in Eq. (4) to calculate the matrix elements for processes like
$B\rightarrow D^*(2010) +\ell+\nu$.
Similar formulas have been derived for $1^-\rightarrow 0^-$ and $1^-\rightarrow
1^-$ transitions, and for processes involving other hadrons containing one or
more heavy quarks.

\twohead{High Precision Electroweak Physics}

  Due largely to the great recent success of LEP,
electroweak physics has become closer and closer in spirit to quantum
electrodynamics, as a branch of physics where high precision is expected.  As
Rubbia said this morning,  we now have three
parameters in  electroweak physics that we know with
quite high precision:
the Fermi coupling constant $G_F$ of beta decay (known from muon decay); the
fine structure constant $\alpha$; and the mass of the Z, measured at LEP.
These three constants (along with masses and mixing angles for whatever quarks
or leptons are involved in a given process) are all you need to calculate any
desired matrix elements in the electroweak theory in tree approximation.
Beyond the tree approximation one needs to know other parameters, like the top
quark and
the Higgs mass, but since these enter in loops observable quantities are less
sensitive to them.  As discussed by Rolandi, with our knowledge of $G_F$,
$\alpha$, and $m_Z$ we are now able to use the electroweak theory to do high
precision calculations of other  measured quantities, such as the W mass, the Z
leptonic width $\Gamma_{\ell}$, and
$[\sin^2 \theta]_Z$ (measured from the forward-backward asymmetry in
$e^++e^-\rightarrow\ell^++\ell^-$ at the Z peak), aside from a weak dependence
on the top and the Higgs mass.   These predictions agree with existing data for
a top quark mass between 130 and
170 GeV (to 68 percent confidence) and a Higgs mass which is essentially
unconstrained, allowed to be anywhere from 50 Gev to 1 TeV.

Satisfactory as this situation is, we can be pretty sure that it will ``soon''
(i. e., within a few years) be radically improved.  We can count on a further
improvement in the accuracy with which $m_W$, $\Gamma_{\ell}$, and $[\sin^2
\theta]_Z$ are known, especially through more accurate measurements of $m_W$ at
LEP2 and the Tevatron Collider, and perhaps also through an improvement in the
measurement
of $[\sin^2 \theta]_Z$ at the SLD.  But the real change will come when the top
quark is discovered and its mass is measured, presumably within a few years at
the Tevatron Collider.  At that point we shall find ourselves in the position
of having a critical
test of the simplest (one scalar doublet) version of the electroweak theory,
and if that test is failed, of being able to say something about what new
physics must be added to this model.

In the last few years it has become customary to
parameterize the new physics that may enter in the electroweak theory in terms
of what are called oblique radiative
corrections.  Parameters like the  Z mass, the Z leptonic width, etc., would be
affected by
the top and Higgs mass, as well as by most kinds of new physics that could be
added to the minimal electroweak theory, mostly through the $2\times 2$ matrix
vacuum polarization of the Z and $\gamma$.  (For instance,   Higgs scalars do
not interact very much with $u$ or $d$ quarks or electrons or muons because
these quarks and leptons are so light.)  Further, as
Peskin explained, because experiments are still mostly done at energies that
are rather
small compared to what we think are the energy scales of new physics, the
vacuum polarization can be parameterized by the coefficients of just the first
few terms in a Taylor expansion.  The most important  of these coefficients
are called $S$ and $T$.  High precision measurements of $m_W$, $\Gamma_{\ell}$,
and $[\sin^2\theta]_Z$ define narrow strips of allowed values in the $S - T$
plane, all running with various slopes from $S<0$, $T<0$ to $S>0$, $T>0$.  For
a given top quark mass (measured, say, to an accuracy of 10 GeV) and a Higgs
mass between 50 GeV and 1 TeV, the
minimal standard model defines a short wedge in the $S - T$ plane, running
roughly transverse to these strips.   [These strips and wedges are shown in
Figure 1.  The strips in this figure are centered on values of $S$ and $T$
derived from the current experimental values of $m_W$, $\Gamma_{\ell}$, and
$[\sin^2 \theta]_Z$.]

There are two possibilities for what will then be found.
\begin{itemize}
\item
If the wedge intersects all the strips then, depending on where along the wedge
the intersection is, we will have a good rough estimate of the Higgs mass in
the minimal standard model.  (It is often said that observables are only very
weakly dependent on the Higgs mass in the minimal standard model, but that is
true only if the top quark mass is not known; predictions for a low Higgs mass
and a low top quark
mass resemble those for a high Higgs mass and a high top quark mass.)  {\em If}
the experimentally allowed ranges of $m_W$, $\Gamma_{\ell}$, and $[\sin^2
\theta]_Z$ continue to be centered on their present values, then this
intersection would indicate a relatively high Higgs mass for $m_t=160\pm 10$
GeV, and  a relatively low Higgs mass for $m_t=120\pm 10$ GeV.

\item
On the other hand, if the wedge doesn't intersect the strips, we will have a
good clue as to what new physics must be added to the minimal electroweak
theory.  For instance, technicolor theories
tend to  move the wedges to larger values of $S$.

\end{itemize}

All this goes to underline the extreme importance of finding the top quark.
\pagebreak

\twohead{Effective Field Theories}

My third topic is not new,  but has lately become part of
the common language of elementary particle physics.  I think that I have heard
the words ``effective
field theory" or ``effective Lagrangian" a hundred times in as many different
contexts at this meeting.
Leutwyler went into effective field theories in some detail, and  Peskin
applied them in discussing oblique radiative corrections.  I remember that
after
Leutwyler's talk, there was a question from the audience asking how we know
that it is legitimate to use these
effective field theories in describing the real world, and then someone else
came up to me in the lobby and asked the
same question.  So although this topic is not by any means new, I would like to
take a few minutes to explain what we are doing when we use an effective field
theory

For this purpose I would like to use an example that is not usually discussed
in terms of
effective field theories,  although I think it's the first example of an
effective
field theory in the literature.  It goes back to the 1930's, when theorists
like Heisenberg were calculating the scattering of light by
light as a somewhat academic application of the new quantum electrodynamics.
In 1936 H.Euler showed that the results for photon-photon scattering amplitudes
at photon energies $\omega\ll m_e$, which are obtained in quantum
electrodynamics
from what would today be called an electron loop graph,
could be summarized as a lowest-order perturbation theory result obtained from
the Lagrangian density:
\pagebreak
\beqra
&&{\cal L}_{\rm eff}=\haf\left(\vec{E}{}^2-\vec{B}{}^2\right)\nonumber\\
&& + \frac{e^4}{360\pi^2m_e^4}
\left[\left(\vec{E}{}^2-\vec{B}{}^2\right)^2+7\left(\vec{E}
\cdot\vec{B}\right)^2\right]\;.
\eeqra
In modern terms, we would say that in order to study processes at energies much
less than $m_e$, we `integrate out' the electron, replacing the Lagrangian of
quantum electrodynamics with the effective Lagrangian (9).

This historical example provides lessons that help us to understand modern
effective field theory:
\begin{enumerate}
\item
We do not really need perturbation theory or even quantum electrodynamics to
understand the general form of Eq. (9).  Gauge and Lorentz invariance tell us
that the most general possible Lagrangian for the electromagnetic field is of
the form
\beqra
&&{\cal L}_{\rm eff}=\haf\left(\vec{E}{}^2-\vec{B}{}^2\right) \nonumber\\&& +
c_1\left(\vec{E}{}^2-\vec{B}{}^2\right)^2
+\,c_2\left(\vec{E}
\cdot\vec{B}\right)^2\nonumber\\&& + c_3 \left(\vec{E}\cdot
\vec{B}\right)\Box\left(\vec{E}\cdot
\vec{B}\right) + \cdots\;,
\eeqra
where the ellipsis ``$\cdots$'' denotes other terms with  derivatives and/or
more electromagnetic field factors, and $c_1, c_2, c_3$, etc., are unknown
constants.
[The factor in the first term is made to be equal to $\haf$ by a canonical
normalization of the fields.]
Dimensional analysis tells us that $c_1$ and $c_2$ have dimensionality
$[mass]^{-4}$, while $c_3$ and all other coefficients have dimensionalities
$[mass]^{-n}$, with $n\geq 6$.  If (10) is obtained by integrating
out `heavy' charged particles  (like the electron) with some typical mass $M$,
then $c_1$ and $c_2$ will be proportional to $M^{-4}$, while $c_3$ and all
other coefficients will be proportional to $M^{-6}$ or higher powers of $1/M$.
If we calculate photon scattering  amplitudes  at photon energies $\omega \ll
M$, then the result will be dominated by the Born approximation contribution of
just the $c_1$ and $c_2$ terms in (10), because all other interactions
and higher order graphs will be suppressed by two or more powers of $\omega/M$.
 [Of course one must go back to quantum electrodynamics to derive the specific
values of the coefficients in Eq. (9), though factors like $e^4$
and $1/8\pi^2$ can be obtained by simply counting vertices and loops in the
graphs from which the effective Lagrangian is calculated.]

\item
The effective Lagrangian (10) is more than an {\em aide memoire} for
photon-photon scattering amplitudes.  It is not hard to show that the dominant
terms in the amplitudes for other low-energy photon reactions like
$\gamma+\gamma\rightarrow\gamma+\gamma+\gamma+\gamma$
are given by using the $c_1$ and $c_2$ terms in (10) in the tree approximation.
 Perhaps more surprisingly, we can regard this effective Lagrangian as the
basis of a  legitimate quantum field theory, calculating corrections to any
process of higher order in $\omega/M$ by
including photon  loop as well as tree diagrams generated by (10).  This
effective field theory is of course non-renormalizable, but the divergences in
these loop graphs are cancelled by renormalization of the coefficients in (10).
Non-renormalizable theories are just as renormalizable as renormalizable
theories; the only difference is that we have to deal with an infinite number
of interaction
types.
You might suppose that the intrusion of unknown constants from the higher terms
in (10) means that this effective theory has lost all predictive power, but
that's not true at all.  For example, the one-loop contribution to
the scattering of light by light gives a term of order $(\omega/M)^8\log
\omega$ with a coefficient given by a known quadratic expression in $c_1$ and
$c_2$, as well as polynomial terms involving the coefficients of higher terms
in (10).

\item
The use of perturbation theory with the Lagrangian (10) does not depend on the
fact that the fine structure constant is small (though that helps);
perturbation theory could be used even if the underlying theory, quantum
electrodynamics, were a strongly interacting theory.  This is because each loop
in the effective field
theory introduces additional factors of $\omega/M$ into the amplitude for any
given process; perturbation theory here is an expansion in powers of
$\omega/M$, as well as $e^2/8\pi^2$.  This of course is why Euler was able to
calculate the leading terms in the photon-photon scattering amplitude by using
(9) in lowest order.   If $e^2/8\pi^2$ were of order unity we would not be able
to derive Eq. (10) from an underlying theory, and we would not be able to
calculate
values for the constants $c_1$, $c_2$, etc., but we could still do perturbation
theory with (10) as our effective Lagrangian.

\item
The crucial feature of the effective Lagrangian (10) that allows us to use it
to generate an expansion for amplitudes in powers of energy is that it involves
only
non-renormalizable interactions, with coupling constants that necessarily have
the dimensions of negative powers of mass.  In pure electrodynamics this is a
consequence of
gauge invariance; renormalizable interactions would be quartic polynomials in
the vector potential $A^\mu$ without derivatives, and these would not be gauge
invariant.

\end{enumerate}

The last remark suggests that we should expect to be able to use effective
Lagrangians to do calculations at low energies whenever there are symmetries
(like gauge invariance in the above example) that rule out renormalizable
interactions.
There are other such symmetries.  One is general covariance in the theory of
gravitation (with no cosmological constant.)  Another well-known example is the
broken chiral $SU(2)\otimes SU(2)$ symmetry of quantum chromodynamics with
massless $u$ and $d$ quarks.  This requires that the pion, the Goldstone boson
of this broken symmetry, enters into the Lagrangian only in
non-renormalizable couplings involving derivatives of pion fields, such as the
leading term in the purely pionic Lagrangian:
\beq
{\cal L}_{\rm eff} = - \frac{\pa_\mu\vec{\pi}\cdot
\pa^\mu\vec{\pi}}{(1+\vec{\pi}{}^2/F_\pi^2)^2}\;.
\eeq
Effective Lagrangians were introduced into modern particle physics in the 1960s
in this context (though we did not know we were doing quantum chromodynamics
then.)  One recent development in this area is the extension of the effective
Lagrangian method to not only
the interactions of pions with each other and with single nucleons, but also to
the interactions of pions and several nucleons - in other words to the
problems of nuclear force and pion scattering on nuclei.  The results obtained
can be summarized
by saying that the chiral Lagrangian approach turns out to justify
approximations (such as assuming the dominance of two-body interactions) that
have been used for many years by nuclear physicists, though not knowing about
the chiral
Lagrangian approach it was not quite fair of them to use these approximations.

The heavy quark symmetries discussed earlier have been incorporated into an
Effective Heavy Quark Field Theory, and this has been combined with the soft
pion theory based on broken chiral symmetries.  Effective Lagrangian techniques
are also being
used today to study not only the ordinary strong interactions, but also
technicolor, a hypothetical extra strong force that has been considered as a
possible source of the breaking of the $SU(2)\otimes U(1)$ electroweak
symmetry.
It is easy to invent technicolor models that have an accidental $SU(2)\otimes
SU(2)$
symmetry containing the $SU(2)\otimes U(1)$ symmetry of the electroweak
interactions, and
it's natural to assume that it is spontaneously broken to a sort of isospin
symmetry called
custodial $SU(2)$, which preserves the usual relation between $W$ and $Z$
masses.  This symmetry breakdown of course entails Goldstone bosons
that provide the longitudinal parts of the $W$ and the $Z$,  and
in most cases additional Goldstone bosons called techni-pions.  The effective
Lagrangian of these Goldstone bosons is very much like Eq. (11), with a
structure dictated by the broken and unbroken symmetries and by the nature of
the degrees of freedom.  As discussed here by Chanowitz, effective Lagrangians
of this sort are actively  being used to study  the possibilities of finding
signs of
technicolor at accelerators like the SSC.

This brings us back to the question that was asked after
Leutwyler's talk: how do we know that it is legitimate to use these effective
Lagrangians?  The derivation of a Lagrangian like (11) is not just a matter of
`integrating out' heavy degrees of freedom in quantum chromodynamics, in the
way that Euler integrated out the electron to obtain (9), because unlike the
electron in QED, the
pion does not appear as a field in the QCD Lagrangian.  True, the breaking of
the $SU(2)\otimes SU(2)$ symmetry implies that there must be a pion particle,
and dictates the structure of any field theory that describes this pion, but
how do we know that this composite pion can be described by a field theory at
all?  The answer seems to be that any quantum theory that satisfies Lorentz
invariance plus a
technical requirement called cluster decomposition plus unitarity will always
at sufficiently low energy look like a quantum field theory.  Quantum field
theory is the only way, we
believe, of reconciling these fundamental requirements.  [The clause ``at
sufficiently low energy" is inserted so that this statement will apply also to
string theories, which contain infinite numbers of increasingly heavy particle
types, and therefore do not look like quantum field theories at energies
comparable to the string mass scale.]  Furthermore, the theory of low energy
pions must be a quantum field theory
described by a broken $SU(2)\otimes SU(2)$ symmetry.  The use of Eq. (11) is
justified not because we derive it from a field theory like QCD, but because it
is the most general chiral invariant theory of pions, aside from terms with
more derivatives whose effects are suppressed at energies $E$ much less than a
QCD scale of order $m_\rho$ by two or more factors of $E/m_\rho$.

The same thought occurs to us with somewhat chilling overtones with regard
to the Standard Model.  We believe that there is some unknown physics at really
high energies, roughly of the order of
the Planck mass, with unbroken symmetries like $
SU(3)\otimes SU(2)\otimes U(1)$ (and perhaps supersymmetry) that keep the
particles of the standard model massless before the breaking of $ SU(2) \otimes
U(1)$ (and supersymmetry).  The most general theory we could expect to find at
low
energies is simply the most general quantum field theory satisfying these
symmetries.  This is of course the standard model, supplemented with
non-renormalizable terms whose effects at energy $E$ are suppressed by powers
of $E/m_{\rm Planck}$.  The success of the standard model tells us nothing
about whether the underlying theory that describes physics at the Planck scale
is a quantum field theory.    In a sense this represents the revenge of
S-matrix
theory,  because we now believe that the field theories of which we are so
proud, quantum electrodynamics, quantum chromodynamics, even general
relativity for that matter, are not what they are because they are truly
fundamental field theories, but simply because they are the only way of
satisfying the requirements of symmetries and
S-matrix theory.  Quantum field theory as we use it now is nothing but
S-matrix theory made practical.  Our best candidate for an underlying theory at
the Planck scale is in fact not a quantum field theory in the ordinary sense,
which brings me to my next topic.

\twohead{Superstring Theory}

Superstring theories have been studied for over a decade as candidates for a
fundamental theory in particle physics.  The implications of superstring
theories were discussed in the rapporteur talks by Alvarez Gaum\'e and Peskin,
and reviewed here in detail in a parallel session talk by Dine; in this summary
I will just make some general remarks about the current state of the theory,
with emphasis on recent work on coupling constant unification in these
theories.

  One change in the last few years has been a movement away from thinking of
superstrings as existing in ten or twenty-six dimensions, of which all but the
four dimensions of ordinary spacetime somehow become compactified, and toward
formulating superstring theories from the beginning
in four spacetime dimensions.  Not that the previous theories are wrong ---
it's just that if one starts with
a ten dimensional superstring theory and then supposes that  six of these
dimensions
become compactified, what you get is the same as if you had started with a
four dimensional superstring theory of an appropriate type.  So  the
four-dimensional approach gives a more general approach to theories that have a
chance to describe nature.

A string as it moves through space sweeps out a two dimensional surface, so the
study of strings is the study of quantum field theories in two dimensions, with
the
spacetime coordinates along with other degrees of freedom appearing as fields
in these theories.  The two surface coordinates in these field theories may be
represented as a single
complex variable $z$, but with the complex plane given an arbitrary topology,
one that becomes more and more complicated as we go to higher and
higher order in perturbation theory.

The action for this two-dimensional theory is a functional of the four
spacetime coordinates $x^\mu(z,z^*)$, plus
[in the popular `heterotic' superstring theories] an equal number of spinor
coordinates $\psi^\mu(z)$, plus a number of other field variables that are
needed to satisfy certain constraints.  These constraints arise ultimately from
the fact that the spacetime metric has $g_{00}=-1$.
This would destroy the positive definiteness of the quantum field theory of the
coordinates $x^\mu(z,z^*)$ if it were not for a symmetry known as conformal
invariance, which allows us to transform away vibrations of the string into the
timelike direction.  But conformal invariance in a theory with just
four $x^\mu(z,z^*)$'s or four $x^\mu(z,z^*)$'s and
four $\psi^\mu(z)$'s is violated by quantum mechanical anomalies similar to the
triangle anomaly in QCD, and these anomalies must be cancelled by the other
fields added to the action.

Each possible equilibrium state of ordinary fields in four dimensions
corresponds to a different two-dimensional conformal field theory, and the
vacuum expectation values of the four-dimensional fields correspond to
parameters in the conformally invariant action.    For instance, there is a
term in the string action that (after a suitable
choice of coordinates in the complex plane) is of the form:
\beqra
&& I_{\rm quad}[x]
 = -\haf\int\!\!\!\int dz\,dz^*\; g_{\mu\nu}(x(z,z^*)) \nonumber\\
&& \quad\times\;\frac{\pa x^\mu(z,z^*)}{\pa z}
    \frac{\pa x^\nu(z,z^*)}{\pa z^*} \;,
\eeqra
where $g_{\mu\nu}(x)$ is the gravitational field, which is required by
conformal invariance to satisfy the Einstein field equations.

There are some useful general results that apply to this whole class of
theories in
the tree approximation, where the complex plane is not given any
 topological complications.  One result is a formula for the string mass, a
parameter related to the string
tension and also
to the slope $\alpha'$  of the Regge trajectories on which the excited states
of the string lie, and which can most simply be regarded as the mass of the
first excited state of
a vibrating string.  It is given in terms of the string coupling constant
(which determines the magnitude of higher order effects) and  the Planck
mass by
\beq
M_{\rm string}\equiv \frac{2}{\sqrt{\alpha'}}=\frac{g_{\rm string}M_{\rm
Planck}}{\sqrt{8\pi}}\;.
\eeq
Also, if physics at energy scales below $M_{\rm string}$ is described by an
$SU(3)\otimes SU(2) \otimes U(1)$ gauge field theory, then the gauge coupling
constants at a renormalization scale $\mu$ near $M_{\rm string}$ approach
values satisfying the relation:
\beq
k_{SU(3)}\,g^2_{SU(3)}=k_{SU(2)}\,g^2_{SU(2)}
=k_{U(1)}\,g^2_{U(1)}=g_{\rm string}^2\;.
\eeq
The constants $k$ are known in the trade as the `levels' of the Kac-Moody
algebra from which the gauge symmetries are derived, but the important thing is
that $k_{SU(3)}$ and $k_{SU(2)}$ are in general integers, in most conformal
field theories just $+1$.  Also, $k_{U(1)}$ takes discrete values (values that
are not changed by continuous changes in the conformal field theory), and in a
large class of interesting models
takes the familiar  value of 5/3.  In other words the relations among gauge
couplings, that until recently have been understood in terms of the `grand
unification' of $SU(3)\otimes SU(2) \otimes U(1)$ in some simple group, occur
naturally in string theory without any need for
grand unification.  In fact, string theory gives a better explanation of these
coupling constant relations than grand unification because in grand unified
theories the presence of color triplet partners of
the Higgs doublet makes it particularly difficult to understand the disparity
between the electroweak scale and the grand unification scale, while such GUT
partners are absent in string theories with  $k_{SU(3)}=k_{SU(2)}=1 $.

Eq. (14) applies to gauge couplings defined at a renormalization scale that is
roughly of the order of the string mass scale (13).  For more precise
information about the value of the scale (or scales) where these relations
apply,  one must look to a one loop calculation.    Such calculations  were
described here briefly by Alvarez Gaum\'{e}, and would have
been described by Peskin if time had allowed.  Including
one-loop corrections, the relation (14) is replaced by
\beq
\frac{16\pi^2}{g_a^2(\mu)} =k_a\, \frac{16\pi^2}{g_{\rm string}^2}+b_a\ln
\left(\frac{M_{\rm string}^2}{\mu^2}\right)+\Delta_a\;,
\eeq
where $b_a$ are the familiar constants appearing in the
one-loop renormalization group equations [$\mu dg_a/d\mu=b_ag_a^3/16\pi^2$] and
$\Delta_a$ are the so-called threshold terms.  In string theory as in field
theory, the threshold terms are given by a sum over all species $n$ of heavy
particles that are integrated out to obtain the effective ``low-energy'' gauge
theory:
\beq
\Delta_a=\sum_n c_{na} \ln \left(\frac{M_n^2}{M_{\rm string}^2}\right)\;,
\eeq
with $M_n$ equal (up to factors of order unity) to the heavy particle masses,
and $c_{na}$ constants
 that depend on the details of the underlying theory.  The $c_{na}$ are of
order unity, and we expect the $M_n$ to be of the order of $M_{\rm string}$, so
the threshold corrections $\Delta_n$ may be expected to be of order unity,
which is why we expect Eq. (14) to apply at a renormalization scale $\mu$ of
order $M_{\rm string}$.

To make a more precise estimate, it is convenient to rewrite Eq. (15) as:
\beq
\frac{16\pi^2}{g_a^2(\mu)} =k_a\, \frac{16\pi^2}{g_{\rm string}^2}+b_a\ln
\left(\frac{M^2}{\mu^2}\right)+\tilde{\Delta}_a\;,
\eeq
where
\beq
\tilde{\Delta}_a\equiv \sum_n c_{na} \ln \left(\frac{M_n^2}{M_{\rm
string}^2}\right)+b_a\ln\left(\frac{M_{\rm string}^2}{M^2}\right)\;,
\eeq
and $M$ is an arbitrary mass.  The point is to try to choose $M$ to minimize
the typical values of $\tilde{\Delta}_a$, in which case according to (17) $M$
is the renormalization scale at which the couplings most closely satisfy (14).
This has been done by a number of authors; what seems to be the best estimate
is:
\beq
M=\frac{e^{(1-\gamma)/2}3^{-3/4}}{\sqrt{2\pi}}\,M_{\rm string}=0.216\,M_{\rm
string}\;.
\eeq
Extrapolating experimental values of the couplings to high energy in the
minimal supersymmetric standard model, the couplings $g^2_{SU(3)}/4\pi$,
$g^2_{SU(2)}/4\pi$, and $\frac{5}{3}g^2_{U(1)}/4\pi$  are found to converge to
a common value $1/(26.3\pm 2.1)$ at a renormalization scale about equal to
$2\times 10^{16}$ GeV.  [It has recently become possible to be much more
precise about this, largely because experimental and theoretical advances have
produced a great improvement in our knowledge of $g^2_{SU(3)}$, reported in a
plenary session here by Bethke.]  With $g_{\rm string}^2/4\pi$
taken equal to $1/26.3$, Eq. (13) gives
$M_{\rm string}=1.7\times 10^{18}$ GeV, and Eq. (19) then gives the
renormalization scale for coupling constant unification as $M=3.6\times
10^{17}$ GeV.  So in other words there is about a factor of twenty discrepancy
between the unification scale expected in string theory,
proportional to the Planck mass,  and the value $2\times 10^{16}$ GeV inferred
from `low' energy experiments.  In speaking of this as a discrepancy, I can't
help feeling a sense of unreality.  Here we are, talking about energy scales
about $10^{13}$ to $10^{14}$ times larger than the largest energies that we can
produce in our
accelerators, and worried about a discrepancy of a factor of 20!

Of course, one possible resolution of this discrepancy is that there may be
more to physics at energies below the string scale than the minimal
supersymmetric standard model, and that the couplings have therefore not been
extrapolated correctly to very high energies.  It is not easy to see what could
be added to the minimal supersymmetric standard
model that would preserve the {\em natural} convergence of all three couplings
to a common value, while moving the energy at which the convergence occurs to
higher values.  Alternatively, the resolution of this discrepancy may be that
the threshold corrections are  larger than we
had thought.  From (16), we see that this could happen if some of the particles
that were `integrated out' in deriving the $SU(3)\otimes SU(2) \otimes U(1)$
standard model are actually somewhat lighter than $M_{\rm string}$.  Such would
be the case if physics immediately below the string scale were described by a
Kaluza-Klein theory, with some extra spatial dimensions forming a compact
manifold
with dimensions somewhat larger than $1/M_{\rm string}$, or by a
four-dimensional grand unified theory, with a simple gauge group broken
spontaneously
to $SU(3)\otimes SU(2) \otimes U(1)$ at an energy somewhat smaller than the
string scale.  But it now seems that if Kaluza-Klein or grand unified theories
are to have any application to the real
world, it will only be in a narrow energy range, roughly  from $4\times
10^{17}$ GeV down to $2\times 10^{16}$ GeV.

  These threshold corrections are important in other
ways.  It has been known for some years that  parameters of the conformal field
theory such as the
dilaton field and modular fields can be fixed only by a non-perturbative
dependence of the vacuum energy on these parameters, because in the range of
these parameters where perturbation theory is valid,  one can show that the
vacuum energy has no local minimum.   [The dilaton field is particularly
important, because it is directly
related to the gauge couplings.]  We know that the vacuum energy depends
non-perturbatively on the couplings $g_a^2(M)$; for instance the coupling
$g_a^2(\mu)$ of QCD or some `hidden sector' gauge field becomes strong at a
renormalization scale $\mu_a$ of order
\beq
\mu_a \approx M\, \exp\left(-\frac{8\pi^2 k_a}{g_a^2(M)}\right)\;,
\eeq
and the vacuum energy density contains terms of order $\mu_a^4$.  But (15)
shows that the couplings $g_a^2(M)$ depend on the threshold correction terms
$\Delta_a$, which depend in a calculable way on parameters like the dilaton and
modular parameter fields,  giving the vacuum energy the desired
non-perturbative dependence on these parameters.

  This analysis has been  explored by many authors in the last few years,
and there are now many candidates for  conformal field theories in which by
`discrete fine tuning' (that is, by choosing
specific models out of a list of thousands, but without carefully adjusting
continuous parameters) you can make the string coupling come out
to have the `observed' value, $g_{\rm string}^2/4\pi = 1/26.3$, and you can
also arrange to have supersymmetry broken at a scale that would give the
gravitino a mass of order 1 TeV, needed to produce electroweak symmetry
breaking
with the observed strength.  This seems to me to
represent real progress in the interaction between superstring theory and
physics.

Speaking of progress, there has been progress of another sort lately.  It has
actually become easier to follow the superstring
literature, because many superstring theorists are now working on
quantum gravity or two-dimensional statistical mechanics, and therefore there
is
less to read that is relevant
to particle physics.

Despite all this progress, superstring theory faces a number of important and
difficult problems.  One of them is to identify correctly the source of
non-perturbative effects.  The non-perturbative effects described above are
found by studying the effective `low' energy quantum field theory derived from
string theory rather from string theory itself, but we do not know if these the
only important non-perturbative effects in string theories.    For some years
it has been widely assumed that the way to get
at other,  really stringy, non-perturbative effects is through what's called
string field theory.  String field theory allows for the creation and
annihilation of strings in much the same way that ordinary field theory
describes the creation
and annihilation of particles.
I have never been enthusiastic about string field theory, because  it seemed to
me to take a beautiful new
formalism and make it ugly by trying to make it look like the old formalism of
quantum field theory, but who knows?

I would like to offer a modest suggestion as to where we might look for
specifically stringy non-perturbative effects without developing a string field
theory.  Let's recall  what Feynman diagrams
signal the advent of non-perturbative effects in ordinary quantum field
theory.
Suppose you want to calculate some process like a vacuum polarization in a pure
non-Abelian gauge field theory, at a
momentum $p$ that is very small compared to the string scale
$M_{\rm string}$.  Instead of introducing  running coupling constants, suppose
we perversely continue to expand in the coupling defined at $M_{\rm string}$
[essentially $g_{\rm string}$], as if Gell-Mann and Low had never been born.
We would then find a breakdown of perturbation theory, because small factors of
$g_a(M_{\rm string})^2/8\pi^2$ would be accompanied with large factors of $\ln
(p/M_{\rm string})$.  These large logarithms come from graphs in which internal
massless gauge boson lines approach the mass shell.    Now, in string theory
the Riemann surfaces
that correspond to these diagrams  have handles that are pulled out to long
thin tubes.  But just as in field theory the large logarithms mean that we have
to deal with loops within loops within etc., in string theory at low momentum
we have to deal with thin handles attached to thin handles attached to etc.
The Riemann surfaces from which non-perturbative effects arise, though
infinitely complicated, may like a fractal surface have a self-similarity
property, of looking qualitatively the same at all scales.

I have no idea how to deal with fractal
Riemann surfaces, but it may be  simpler than the task of dealing with the
corresponding non-perturbative effects in quantum
field theory, because as Alvarez Gaum\'e emphasized, in some respects string
theory is much simpler than quantum field theory.  Gauge invariance appears
more naturally, and there are fewer diagrams.  This aspect of string theory has
been exploited as a calculational device in quantum chromodynamics, and may
perhaps allow us to understand stringy non-perturbative effects that go beyond
anything we have been able to understand in quantum field theory.

Perhaps the most fundamental
problem facing string theory is that we still do not know even in principle
what it is that chooses the correct string theory, corresponding to the correct
vacuum.  The wrong answer is that the correct vacuum is the one with
lowest vacuum energy, because we already know plenty of superstring theories
that have negative vacuum
energy.  There is another possible answer supplied by
quantum cosmology.  In recent years the study of wormhole effects has suggested
that the universe is not in a state in which all coupling constants have
definite values, but rather in a quantum mechanical superposition of such
states.  Some relations among coupling constants are fixed by fundamental
principles (such as the fact that the electric charges of the electron and
positron are equal and opposite) and would be the same in all terms in this
superposition,  but any  coupling
constant that {\em can} vary continuously from one theory to another would have
a continuous range of values in the different terms in the state vector of the
universe.  There may not be any such free parameters, in which case the
following discussion is irrelevant, but our failure to formulate any principle
for choosing the vacuum state in string theories suggests that the vacuum
energy may be a free parameter that varies continuously from term to term in
the state vector of the universe.    When the universe becomes large, as it is
now, it begins to look like an incoherent
mixture of these terms, with various probabilities.  In this case, it is only
common sense that scientists who worry about the vacuum energy would have to
find a value in
the fairly  narrow range in which life could arise; all the other terms in the
wave function are there, but there are no scientists to observe them.  The
vacuum energy acts in cosmology like Einstein's cosmological constant.  It
can't be too positive
because then galaxies would never have formed, and it can't be too negative
because then the universe wouldn't live long enough for life to evolve.  On
this basis we can understand in a natural way  why the vacuum
energy is relatively small - some hundred and twenty orders of magnitude
smaller than we would guess (a Planck mass per cubic Planck length) from purely
dimensional considerations.  But these considerations do not tell us that the
vacuum energy (or equivalently, the cosmological constant) is zero, or even
that it is astronomically negligible, but only  that it is
less than about a hundred times the present mass density of the universe.
This is an interesting bound, because both the cosmological missing mass
problem and the cosmic age problem mentioned by Krauss (the fact that age of
globular clusters seems to be larger than some estimates of the age of the
universe) would be alleviated by a positive
cosmological constant corresponding to a vacuum energy roughly ten times the
present mass density of the universe.
It will be interesting to see whether this is the case.
\begin{center}
\vglue-3pt
*\quad *\quad *
\end{center}
\vglue-3pt

This concludes my survey of special topics.  I will now  stick my neck way out,
and try to guess what lies ahead for particle physics.
\begin{itemize}
\item

All the
present experimental challenges to the standard model will disappear.  This
judgment is based in part on having lived through it all before.  Experiments
certainly from time to time have required the incorporation of new features in
the standard model, such as the tau lepton and the bottom quark.  But where new
experimental results have proved indigestible, irreconcilable with the general
framework of the standard model, they have always gone away.  Among these
supposed difficulties were the high-$y$ anomaly, parity conserving
neutral currents, anomalous trimuon events, and second-class currents.  (Some
of us suffer just hearing this list.)  The two outstanding present experimental
difficulties with the standard model are  the tau branching ratios discussed by
Drell and the
seventeen kilovolt neutrino discussed by Robertson.   Drell's talk hints that
the
tau branching ratio problem is beginning to go away (although it certainly
hasn't gone
away yet), and Robertson came out pretty strongly against the reality of the
seventeen kilovolt neutrino.  So it takes no great courage to predict that
these anomalies will go away as well.
\item
My second guess (these are all just guesses) is that the electroweak
symmetry will turn out to be broken by the vacuum expectation values of
elementary scalars that appear in the
effective Lagrangian at accessible energies,
like  the scalar doublet in the original electroweak theory, and that the
hierarchy
problem will be solved by supersymmetry.  I say this for a number of reasons.
As Peskin discussed, the technicolor idea requires awkward extensions to
produce the quark and lepton masses without introducing new difficulties like
flavor changing neutral currents.  Also, as Peskin and Rubbia remarked, some
simple technicolor  theories are already excluded by the high precision
electroweak data discussed above.
 Another reason for this guess is that I find the convergence of the
$SU(3)\otimes SU(2) \otimes U(1)$ couplings in the supersymmetric standard
model very impressive, and this convergence is easily lost if you mess up the
model by adding things like technicolor.  My last reason has to do with the
solar neutrino problem, and is explained below.

\item
I would guess that the solar neutrino deficit is real
and is in fact explained by the MSW effect.  This is in part because of the
present
state of the neutrino experiments.  As discussed  by Krauss and in a parallel
session by Bahcall, it is not possible by adjusting the temperature at the
center of the sun to make the standard solar model fit both the chlorine and
Kamiokande data, but calculations based on the MSW effect can fit everything,
including the data from SAGE and GALLEX.  Furthermore, from a
theorist's point of view the neutrino mass-square difference $\Delta m^2_\nu$
and mixing angle $\theta_\nu$ that are needed to fit this data are very
plausible:  $\theta_v$  is like a typical small mixing angle in the
CKM
matrix, and $\sqrt{\Delta m^2_\nu}$ is a few millivolts, which is just what
would be expected in the simplest extensions of the standard model.  As is
always the case for effective field theories, the Lagrangian of the standard
model must be supplemented with
non-renormalizable terms that are suppressed by powers of some large mass $M$,
such as $10^{16}$ GeV.  The least suppressed non-renormalizable term is a
quartic term of dimension five, involving two factors of both the  Higgs
doublet and the lepton doublets:
\beqra
&&{\cal L}_5= \frac{g_{ij}^2}{M}\left[\left(\begin{array}{c}\phi^0 \\
\phi^+\end{array}\right)\cdot \left(\begin{array}{c}\nu_i \\
\ell_i\end{array}\right)\right]\nonumber\\
&&\times\left[\left(\begin{array}{c}\phi^0 \\
\phi^+\end{array}\right)\cdot \left(\begin{array}{c}\nu_j \\
\ell_j\end{array}\right)\right]\;,
\eeqra
that after electroweak symmetry breaking yields a Majorana neutrino mass
matrix:
\beq
m^2_{ij}=g_{ij}\lag \phi^0 \rag^2_{\rm vac}/M\;.
\eeq
If we take the couplings $g_{ij}$ to be of the order of the product of the
Yukawa couplings of the $i$'th and $j$'th lepton doublets to the scalar
doublets, then the largest neutrino mass is of the order of $m_{\rm top}^2/M$,
which for $M\approx 10^{16}$ GeV is indeed a few millivolts.
But this attractive picture of the origin of the neutrino masses needed in the
MSW effect is only possible if the scalars are elementary.  If in  a
technicolor
theory  you try to construct lepton number violating non-renormalizable
interactions, you must replace the scalar doublet with some sort of bilinear
function of techniquark fields, but then (21) is replaced with with an operator
of very high dimension, which is strongly suppressed by many factors of $1/M$.
Of course, a neutrino mass of a few millivolts might arise  from all sorts of
possible  new physics, like lepton symmetry breaking at the
technicolor scale, but there would be no particular reason to expect millivolt
neutrino masses.  This gives a special importance to studies of solar neutrinos
and neutrino oscillations, but the pace of these experiments is unfortunately
very slow, a little like real time studies of continental
drift.  In the next few years we can look forward to SAGE and
GALLEX
being calibrated with artificial megacurie neutrino sources, to super
Kamiokande coming into being, and to the start of the SNO experiment
in Canada.  All of these are important, but I want to emphasize that the SNO
experiment offers the possibility of measuring a process $\nu+d\rightarrow
\nu+p+n$ that arises solely from weak neutral currents and should therefore be
unaffected by the MSW effect.  If the MSW effect is indeed solely responsible
for the observed neutrino deficit, then this neutral current process should be
observed at precisely the rate predicted by the standard solar model,  which
may be our best way of reassuring ourselves that the sun is really well
understood.

\item
We are going to find a great deal of new physics at accessible accelerator
energies.  With supersymmetry invoked to solve the hierarchy problem, we expect
not only sparticles, but also flavor changing neutral current processes that
as Peskin
emphasized are endemic in supersymmetry theories.  Also endemic in
supersymmetry theories are CP violations that go beyond the CKM matrix, and for
this reason it may be that the next exciting thing to come along will be the
discovery of a neutron or atomic or electron electric dipole
moment.  These electric dipole moments were just
briefly mentioned at this conference, but they seem to me to offer one of the
most exciting possibilities for progress in particle
physics.  Experiments here as in solar neutrino physics move very slowly, but I
should mention that there has been a lot of progress lately
in calculating the electric dipole moment of atoms in various models, with
results that are encouraging for future experiments.

\item
The correct theory underlying the standard model is probably a superstring
theory.  So far, our best proof consists of asking what else it could
 be.  Superstring theories may be confirmed (and here I'm saying something that
Peskin especially
wanted me to say) by predictions for the coefficients of soft supersymmetry
 breaking terms in the supersymmetric standard model.  In particular, it has
been recently realized that in
superstring theories it's typical that the lightest superparticles are
gauginos rather than squarks or sleptons.  This brings me to my final rash
remark:
\item
Photinos are the cold dark matter needed in galaxy clusters.
\end{itemize}
\begin{center}
*\quad * \quad *
\end{center}
In sticking my
neck way out
and acting as if we really are beginning to
see the final answer I am going against the conventions of conference
summarycraft.    Michael Peskin showed a cartoon that expresses a more
common view: a complacent physicist working away at a computer terminal,
oblivious to  monsters lurking just behind a wall.  It is usual to end a
conference summary with a remark that of course we expect that we
will discover entirely new physics and that we are very far from anything like
a final answer.  But just because this is the conventional view and has
generally been true in the past does not mean that it is true now.  Michael
mentioned a geographical
metaphor: Columbus expecting to sail straight to the Indies, and not
realizing there was something equally interesting in the way, namely America.
[By the way, that geographical metaphor was used at a breakfast meeting in 1987
in convincing the Secretary
of Energy to support the SSC.]  But maybe a different geographical
metaphor is more to the point.  Imagine polar explorers sitting in
the Travellers Club in the late nineteenth century, saying over their port,
"You know, no
matter how far north one goes, there's always plenty of sea and ice left
further
north.  No matter how far north we go, we shall never get to the North Pole."
Well, eventually explorers did get to the
North Pole.  We too may eventually get to our destination, to a final theory,
and possibly sooner rather than later.

\twohead{Acknowledgement}

I wish to thank John Bahcall, Michael Dine, Lance Dixon, Willy Fischler, Howard
Georgi, Nathan Isgur, Vadim Kaplunovsky, Michael Peskin, and Joe Polchinski for
helpful conversations regarding topics covered in this report.

\end{document}